\documentclass[]{raa}            
\usepackage{graphicx,times}
\usepackage{natbib}
\input{epsf.sty}                        
\input{psfig.sty}
\providecommand{\bm}[1]{\mbox{\boldmath$#1$\unboldmath}}

\makeatletter

\newcommand{\Rmnum}[1]{\expandafter\@slowromancap\romannumeral #1@}

\begin{document}

\title{Cosmological Constraints on the Undulant Universe}


   \author{Tian Lan
      \inst{1,2}
   \and Yan Gong
      \inst{2,3}
   \and Hao-Yi Wan
      \inst{4}
   \and Tong-Jie Zhang
      \inst{1,5}
   }

   \institute{Department of Astronomy, Beijing Normal University, Beijing, 100875, China;
        \and
             National Astronomical Observatories, Chinese Academy of Sciences, Beijing 100012,
             China;
        \and
             Graduate School of Chinese Academy of Sciences, Beijing 100049,
             China;
        \and
             Business Office, Beijing Planetarium, No. 138 Xizhimenwai Street, Beijing 100044,
             China;
        \and
             Center for High Energy Physics, Peking University, Beijing 100871, P. R. China \\{\it tjzhang@bnu.edu.cn}
}

\abstract{ We use the redshift Hubble parameter $H(z)$ data derived from relative galaxy ages, distant type Ia supernovae (SNe Ia), the Baryonic
Acoustic Oscillation (BAO) peak, and the Cosmic Microwave Background (CMB) shift parameter data, to constrain cosmological parameters in the
Undulant Universe. We marginalize the likelihood functions over $h$ by integrating the probability density $P\propto e^{-\chi^2/2}$. By using
the Markov Chain Monte Carlo (MCMC) technique, we obtain the best fitting results and give the confidence regions on the $b-\Omega_{\rm m0}$
plane. Then we compare their constraints. Our results show that the $H(z)$ data play a similar role with the SNe Ia data in cosmological study.
By presenting the independent and joint constraints, we find that the BAO and CMB data play very important roles in breaking the degeneracy
compared with the $H(z)$ and SNe Ia data alone. Combined with the BAO or CMB data, one can improve the constraints remarkably. The SNe Ia data
sets constrain $\Omega_{\rm m0}$ much tighter than the $H(z)$ data sets, but the $H(z)$ data sets constrain $b$ much tighter than the SNe Ia
data sets. All these results show that the Undulant Universe approaches the $\Lambda \rm$CDM model. We expect more $H(z)$ data to constrain
cosmological parameters in future. \keywords{Cosmology -- Observational cosmology -- Dark matter} }

  \authorrunning{Tian Lan, Yan Gong, Hao-Yi Wan \and Tong-Jie Zhang}   
  \titlerunning{Cosmological Constraints  on the Undulant Universe}   
  \maketitle

%
%
\section{Introduction}           
\label{sect:intro} Many cosmological observations
(\cite{Bennett2003}; \cite{Spergel2007}; \cite{Eisenstein2005};
\cite{Kowalski2008a}), such as distant SNe Ia (\cite{Riess1998};
\cite{Perlmutter1999}), the subsequent CMB measurement by
Wilkinson Microwave Anisotropy Probe (WMAP) (\cite{Spergel2003})
and the large scale structure survey by Sloan Digital Sky Survey
(SDSS) (\cite{Tegmark2004a}; \cite{Tegmark2004b}) etc. support
that our Universe is undergoing an accelerated expansion, and
these data have mapped the universe successfully by constraining
cosmological parameters (\cite{Perlmutter1999};
\cite{Tegmark2000}; \cite{Zhu2004}; \cite{Zhan2006};
\cite{Lineweaver1998}; \cite{Efstathiou1999}; \cite{Zhan2005};
\cite{Maartens2006}). Meanwhile, many different observations, such
as the redshift-dependent quantities, are used usually. The
luminosity distance to a particular class of objects, e.g. SNe Ia
and Gamma-Ray Bursts (GRBs) (\cite{Nesseris2004}), the size of the
BAO peak detected in the large-scale correlation function of
luminous red galaxies from SDSS (\cite{Eisenstein2005}) and the
CMB data obtained from the three-year WMAP estimate
(\cite{Wang2006}; \cite{Spergel2007}) are redshift-dependent. The
BAO and CMB data have been combined to constrain the cosmological
parameters widely. Recently, the Hubble expansion rate at
different redshifts can be obtained from the measurement of
relative galaxy ages directly. As a function of redshift $z$, the
$H(z)$ data have been used to test cosmological models
(\cite{Gong2008}; \cite{Yi2007}; \cite{Samushia2006};
\cite{Wei2006}; \cite{zjf2007}; \cite{Wei2007}; \cite{Zhangx2007};
\cite{Danta2007}; \cite{Zhanghs}; \cite{Wu+}; \cite{Wei+}). Using
different data to constrain parameters can provide the consistency
checks between each other, and the combination of different data
can also make the constraints tighter (\cite{lh2008};
\cite{zgb2007}; \cite{xjq2006}). So far, few work has been done in
comparison. We'd like to study the comparison of the SNe Ia,
$H(z)$, BAO and CMB data in this letter. Although lots of
cosmological models, e.g. the Quintessence (\cite{Caldwell1998}),
the brane world (\cite{Deffayet2002}), the Chaplygin Gas
(\cite{Alcaniz2003}) and the holographic dark energy models
(\cite{Ke2005}), are extensively explored to explain the
acceleration of the universe. $\Lambda \rm$CDM model, which is a
standard and popular model, has been studied by Lin et al.
(\cite{Hui2008}). In fact, there are some problems in the standard
model, and this model is not consistent with the real universe.
Scientists proposed a Undulant Universe, which is characterized by
alternating periods of acceleration and deceleration
(\cite{Gabriela2005}). This model can remove the fine tuning
problem (\cite{Vilenkin}; \cite{Jaume1999}; \cite{Bludman2000};
\cite{Garriga2001}; \cite{Stewart2000}), and solve the coincidence
scandal between the observed vacuum energy and the current matter
density, with some details in (\cite{Peebles2003}). The equation
of state (EOS) of the Undulant Universe is
\begin{equation}
\omega(a)=-\cos(\rm b\ln a),\label{eos}
\end{equation} where the dimensionless parameter $b$
controls the frequency of the accelerating epochs. Due to most
inflation models predicting $\Omega_{\rm k}< 10^{-5}$
(\cite{TR1998}), we assume spatial flatness, so the Hubble
parameter is
\begin{eqnarray}
H^{2}(a)=H_{0}^{2}(\Omega_{\rm
m0}a^{-3}+\Omega_{\Lambda0}a^{-3}\exp[\frac{3}{b}\sin(b\ln
a)]),\label{h}
\end{eqnarray}where $\Omega _{\rm m0}$, $\Omega_{\Lambda0}$ are the density
parameter of matter and dark energy, respectively. Furthermore, S.
Nesseris and L. Perivolaropoulos have enhanced the fact that an
oscillating expansion rate ansatz has provided the best fit to the
data among many ansatz by using recent supernova data(\cite{NP2004}). 

In this letter, we concentrate on the comparison of different data
sets on the Undulant Universe. In \S \Rmnum{2}, we simply
introduce these data sets that we use. Then the constraints are
shown in \S \Rmnum{3}. Finally, we show some important conclusions
and discussions in \S \Rmnum{4}.

\section{Observational data}

\subsection{SNe Ia data}
There has been some calibrated SNe Ia data with high confidence in
(\cite{Kowalski2008a}). This data set contains 307 data and the
redshifts of these data span from about 0.01 to 1.75. These data
give luminosity distances $d_{\rm L}(z_{i})$ and the redshifts
$z_{i}$ of the corresponding SNe Ia. In a
Friedmann-Robertson-Walker (FRW) cosmology, considering a flat
universe, the luminosity distance is,

\begin{equation}
d_{\rm L}(z)=\frac{c(1+z)}{H_0}{\mathcal F}(z),
\end{equation}

The function ${\mathcal F}(z)$ is defined as ${\mathcal
F}(z)=\int_0^z{\rm d}z/E(z)$, with
$E(z)=H(z)/H_0=\sqrt{\Omega_{M0}a^{-3}+\Omega_{\Lambda0}a^{-3}\exp[\frac{3}{b}\sin(b\ln
a)]}$. $H_0$ is the Hubble constant. The distance modulus is

\begin{equation}
\mu(z)=\rm m-\rm M=5\log\frac{d_{\rm L}}{10{\rm pc}}=42.39+5\log\
\frac{1+z}{h}{\mathcal F}(z),
\end{equation}

with $h=H_0/100$ km s$^{-1}$ Mpc $^{-1}$, where $\rm m$ and $\rm
M$ are the apparent and absolute magnitudes, respectively.

\subsection{The BAO Data}
As the acoustic oscillations in the relativistic plasma of the
early universe will also be imprinted onto the late-time power
spectrum of the non-relativistic matter (\cite{Eisenstein1998}),
the acoustic signatures in the large-scale clustering of galaxies
yield additional tests for cosmology (\cite{Spergel2003}).
Eisenstein et al. (\cite{Eisenstein2005}) choose a spectroscopic
sample of 46748 luminous red galaxies from SDSS to find the peaks
successfully. These galaxies cover 3816 square degrees and have
been observed at redshifts up to $z=0.47$. The peaks are described
by the model-independent $\mathcal {A}$-parameter. $\mathcal
{A}$-parameter is independent of Hubble constant $H_0$,
\begin{equation}
\mathcal {A}=\frac{\sqrt{\Omega_{\rm
m}}}{z_1}[\frac{z_1}{E(z_1)}\mathcal
{F}^{2}(z_1))]^{1/3},\label{eq3}
\end{equation}
where $z_1=0.35$ is the redshift at which the acoustic scale has
been measured. Eisenstein et al. suggested that the measured value
of the $\mathcal {A}$-parameter is $\mathcal {A}=0.469\pm0.017$
(\cite{Eisenstein2005}).

\subsection{The CMB Data}
The expansion of the Universe has transformed the black body
radiation, which is left over from the Big Bang, into the nearly
isotropic 2.73 K cosmic microwave background (CMB)
(\cite{Bernardis2000}). The whole shift of the CMB angular power
spectrum is determined by the CMB shift parameter $\mathcal {R}$.
The shift parameter perhaps the most model-independent parameter,
is also independent of $H_0$. It can be derived from CMB data
(\cite{Bond1997}; \cite{Odman2003}),
\begin{equation}
\mathcal {R}=\sqrt{\Omega_{\rm m}}\mathcal {F}(z_{\rm
r}),\label{eq6}
\end{equation}
where $z_{\rm r}=1089$ is the redshift of recombination. Using the
Markov Chain Monte Carlo (MCMC) chains from the analysis of the
three-year results of WMAP (\cite{Spergel2007}), Wang \& Mukherjee
compute the CMB shift parameter to get $\mathcal {R}=1.70\pm0.03$
and demonstrate that its measured value is mostly independent of
assumptions about dark energy (\cite{Wang2006}).

\subsection{the $H(z)$ data from relative galaxy ages}
The Hubble parameter $H(z)$ data can be derived from the
derivative of redshift $z$ with respect to the cosmic time $t$,
i.e. ${\rm d} z/{\rm d}t$ (\cite{Jimenez2002}),
\begin{equation}
H(z)=-\frac{1}{1+z}\frac{{\rm d}z}{{\rm d}t}.\label{eq2}
\end{equation}
Therefore, an application of the differential age method to old
elliptical galaxies in the local universe can determine the value
of the current Hubble constant. This is a direct measurement for
$H(z)$ through a determination of ${\rm d}z/{\rm d}t$. Jimenez et
al. (\cite{Jimenez2003}) have applied the method to a $z \leq 0.2$
sample. Paying careful attention to uncertainties in the distance,
systematics, and model uncertainties, they have selected
Monte-Carlo techniques to demonstrate the feasibility by
evaluating the errors. They also use many other different methods
and yield consistent value for the Hubble constant. Hence, the
$H(z)$ data are reliable. With the availability of new galaxy
surveys, it becomes possible to determine $H(z)$ at $z>0$. By
using the differential ages of passively evolving galaxies
determined from the Gemini Deep Deep Survey (GDDS)
(\cite{Abraham2004}) and archival data (\cite{Treu2001};
\cite{Treu2002}; \cite{Nolan2003a}; \cite{Nolan2003b}), Simon et
al. derive a set of $H(z)$ data, which is listed in
Table.\ref{Table 1} (\cite{Simon2005}, see also
\cite{Samushia2006}; \cite{Jimenez2003}). The estimation method in
detail can be found in the work (\cite{Simon2005}). As $z$ has a
relatively wide range, $0.1<z<1.8$, these data are expected to
provide a full-scale description for the dynamical evolution of
our universe. The application of the $H(z)$ data to cosmology can
be referred to (\cite{Simon2005}; \cite{Yi2007};
\cite{Samushia2006}; \cite{Wei2006}; \cite{Hui2008};
\cite{Gong2008}) and so on.

\begin{table}
\small \centering

\begin{minipage}[]{120mm}
\caption{the $H(z)$ data in units of km s$^{-1}$Mpc$^{-1}$
(\cite{Simon2005}) (see \cite{Samushia2006}; \cite{Jimenez2003}
also)}\label{Table 1}\end{minipage}
\tabcolsep 6mm
\begin{tabular}{lcrlcccccccc}
  \hline\noalign{\smallskip}
    $z$(redshift)  & 0.09 & 0.17 & 0.27 & 0.40 & 0.88 & 1.30 & 1.43 & 1.53 & 1.75\\
    $H(z)$    & 69 & 83   & 70   & 87   & 117  & 168  & 177  & 140  & 202\\
    1$\sigma$ error & 12.0 & 8.3 & 14.0 & 17.4 & 23.4 & 13.4 & 14.2 & 14.0 & 40.4\\
  \noalign{\smallskip}\hline
\end{tabular}
\end{table}

\begin{figure*}
\begin{tabular}{cc}

\parbox{58mm}{\psfig{figure=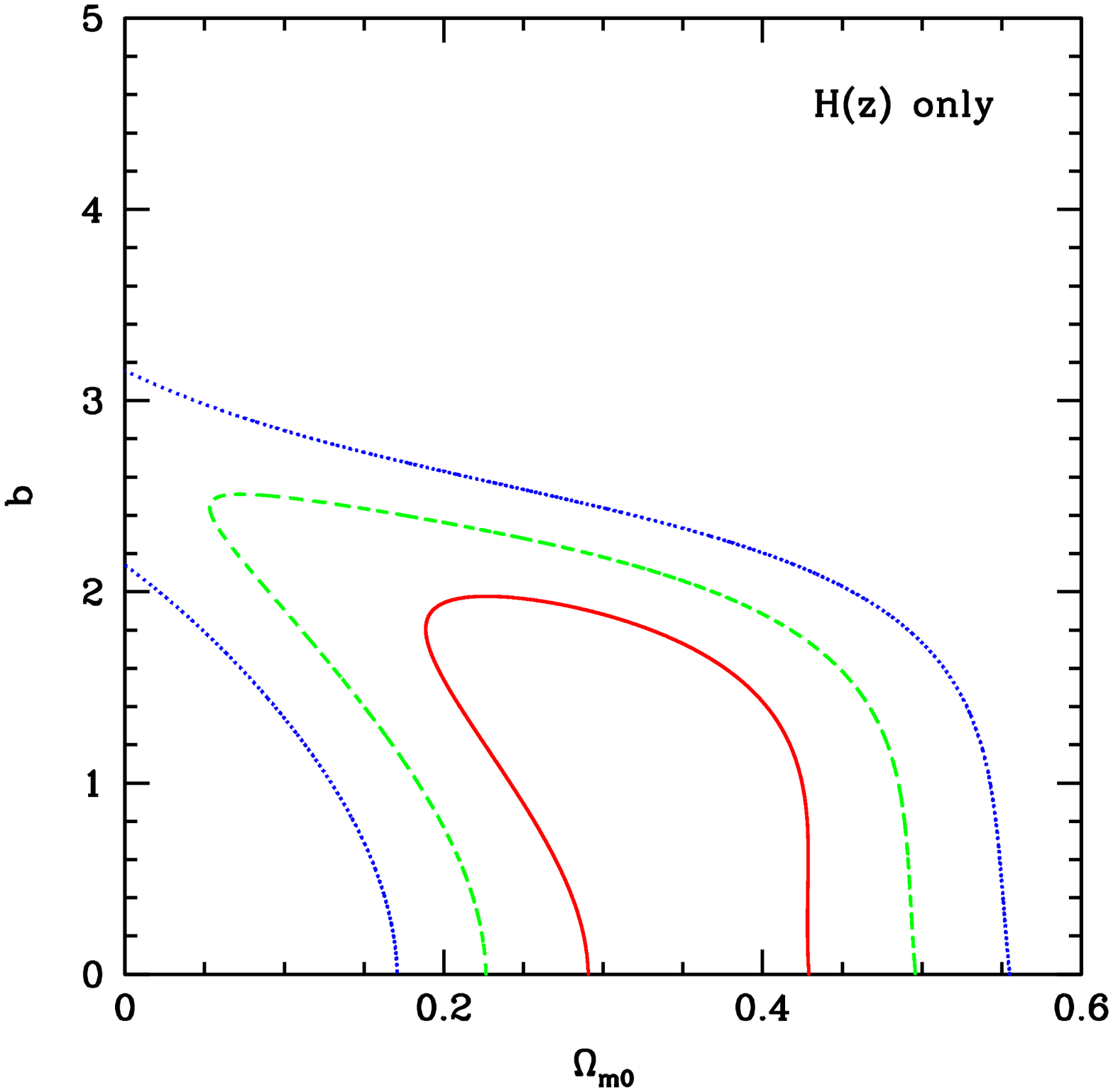,width=3.truein,height=3.truein,angle=0}}\
\parbox{58mm}{\psfig{figure=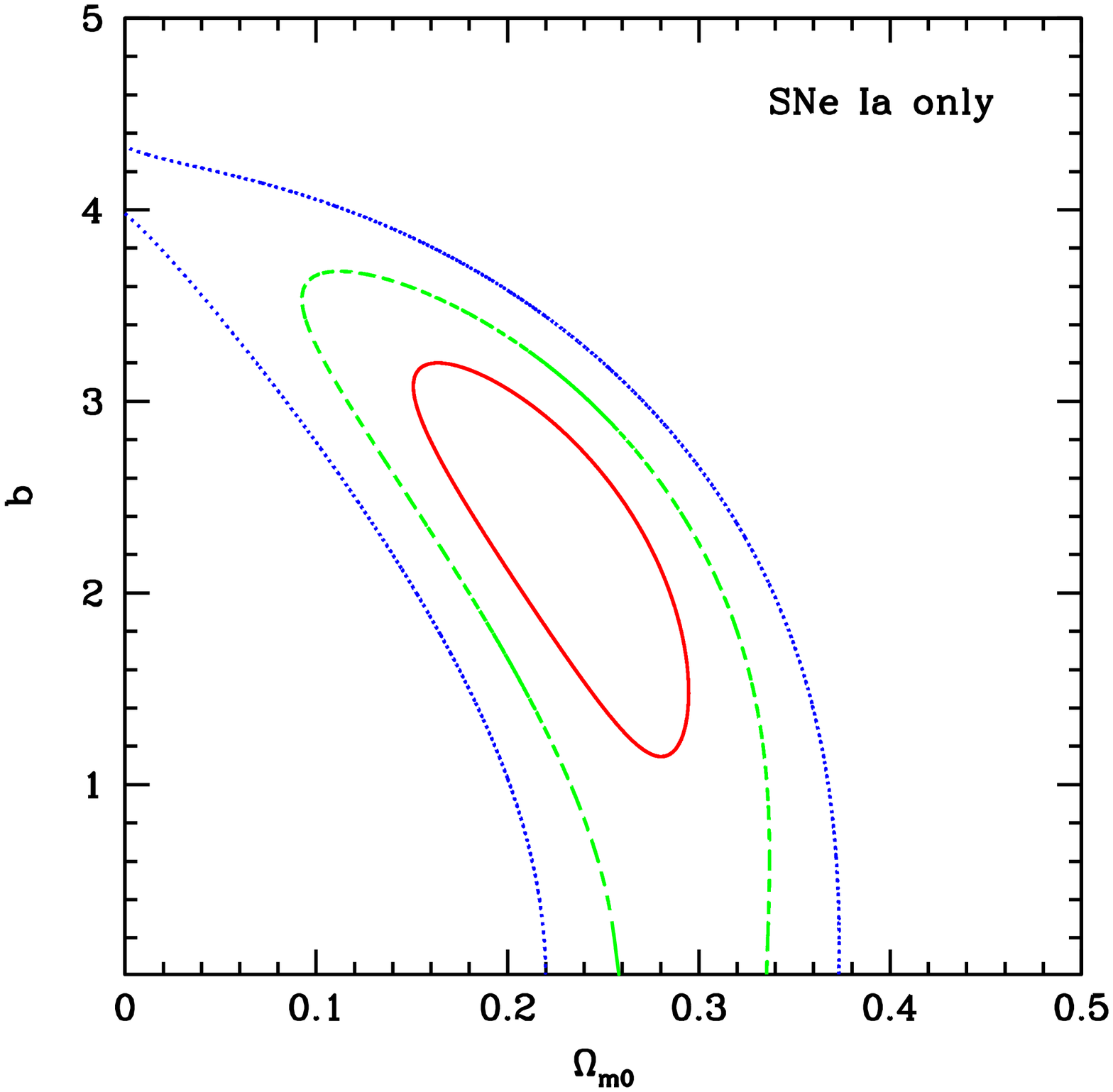,width=3.truein,height=3.truein,angle=0}}\\
\end{tabular}
\caption{The contour maps of $b-\Omega_{\rm m0}$ for different
data sets. The left panel is $H(z)$ only, and the right panel is
SNe Ia only. Confidence regions are at $68.3\%$, $95.4\%$ and
$99.7\%$ levels from inner to outer respectively, for a flat
Undulant Universe without a prior of $h$.}\label{fig.1}
\end{figure*}

\begin{figure*}

\begin{tabular}{cc}

\parbox{58mm}{\psfig{figure=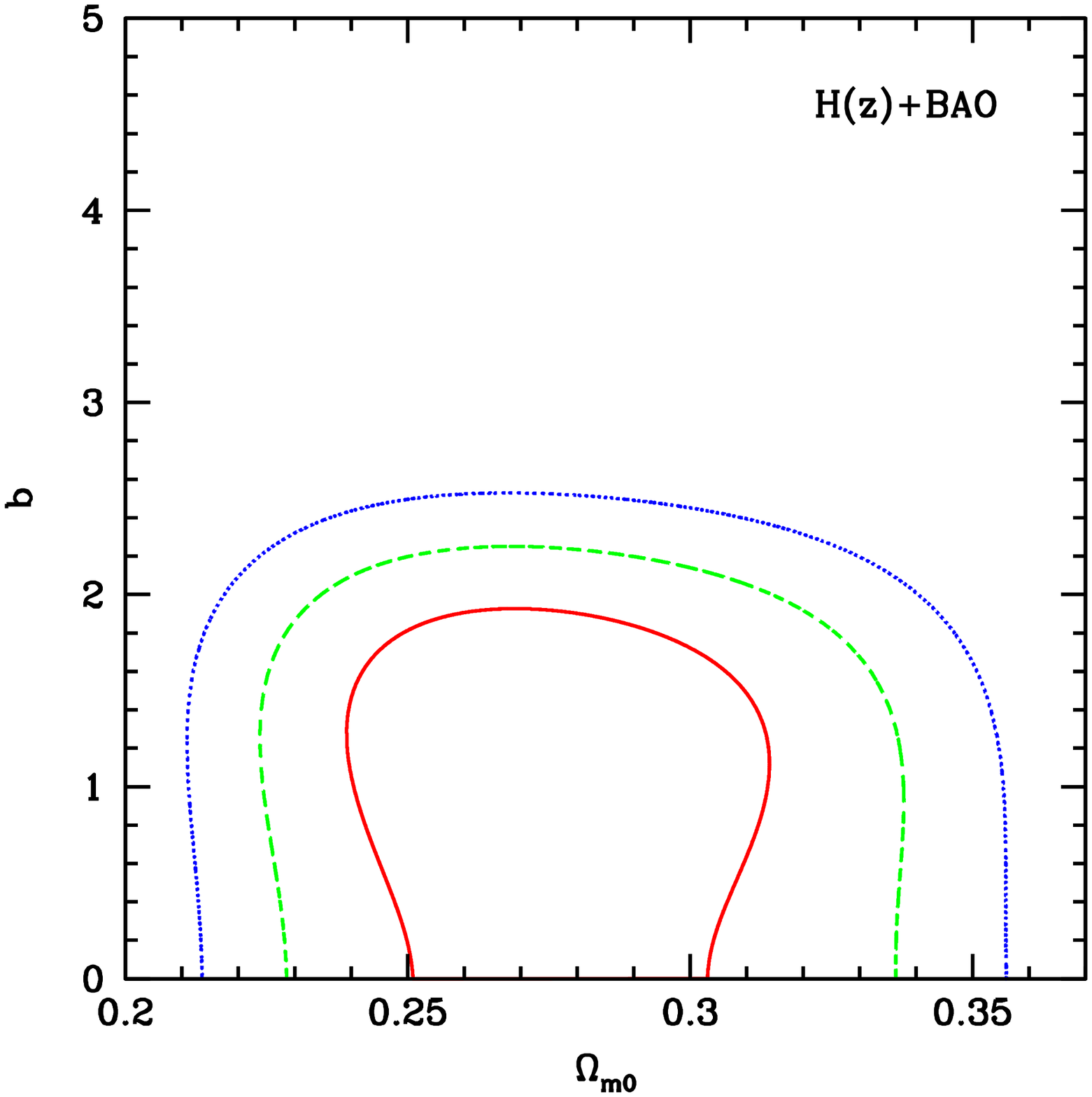,width=3.truein,height=3.truein,angle=0}}\
\parbox{58mm}{\psfig{figure=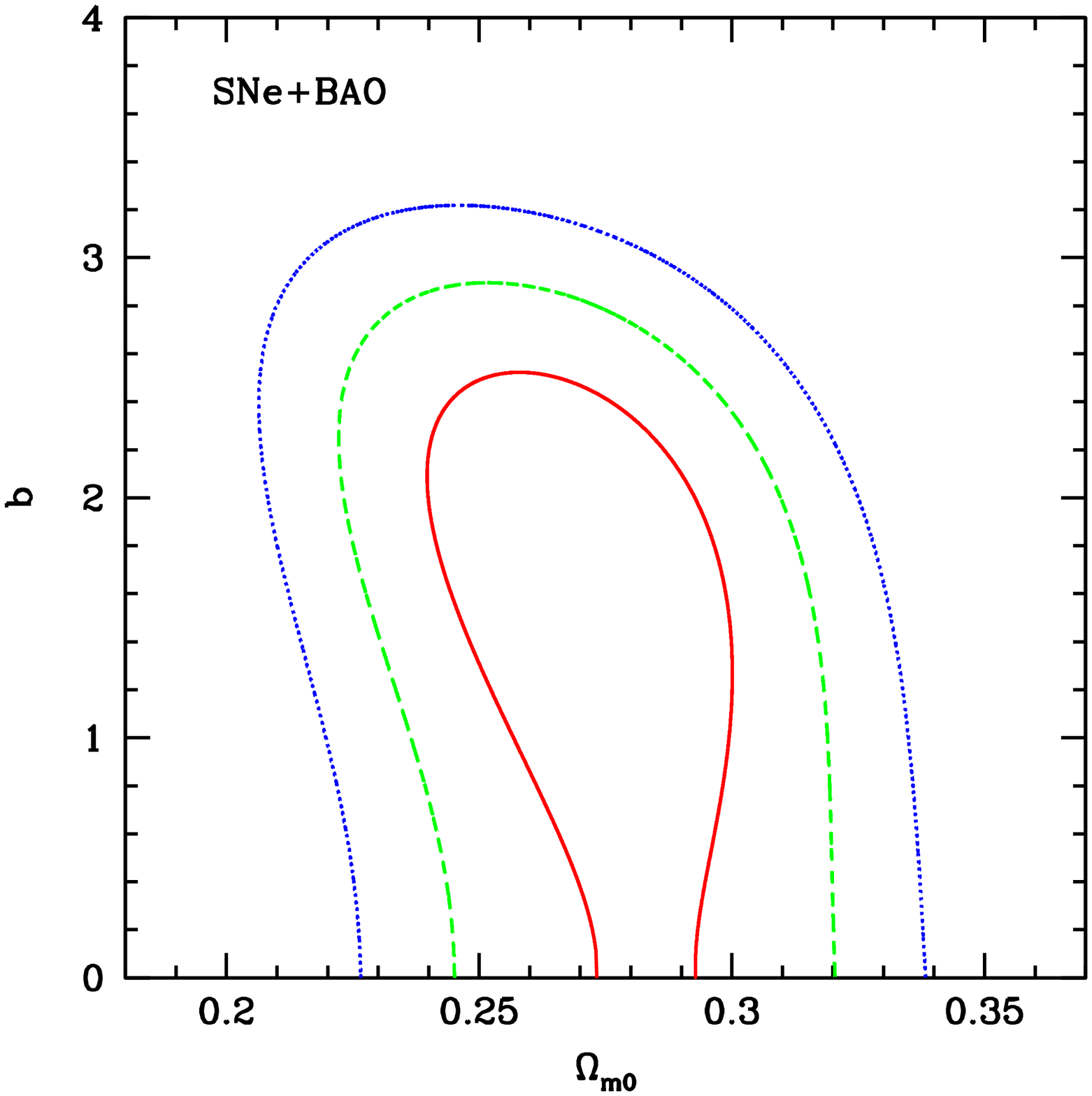,width=3.truein,height=3.truein,angle=0}}\\
\end{tabular}
\caption{The contour maps of $b-\Omega_{\rm m0}$ for different
sets. The left panel is $H(z)$+BAO, and the right panel is SNe
Ia+BAO. Confidence regions are at $68.3\%$, $95.4\%$ and $99.7\%$
levels from inner to outer respectively, for a flat Undulant
Universe without a prior of $h$.}\label{fig.2}
\end{figure*}

\begin{figure*}

\begin{tabular}{cc}

\parbox{58mm}{\psfig{figure=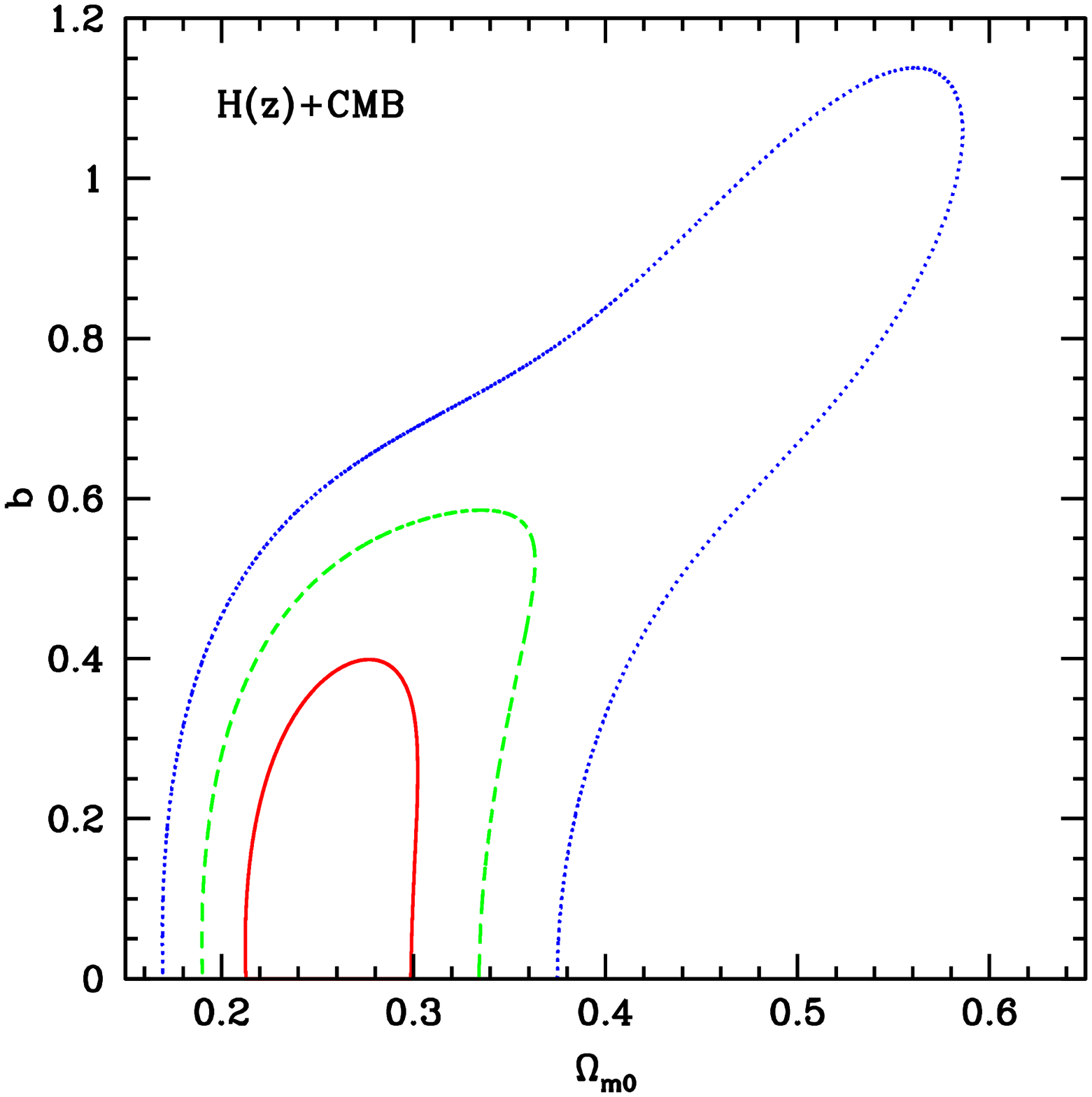,width=3.truein,height=3.truein,angle=0}}\
\parbox{58mm}{\psfig{figure=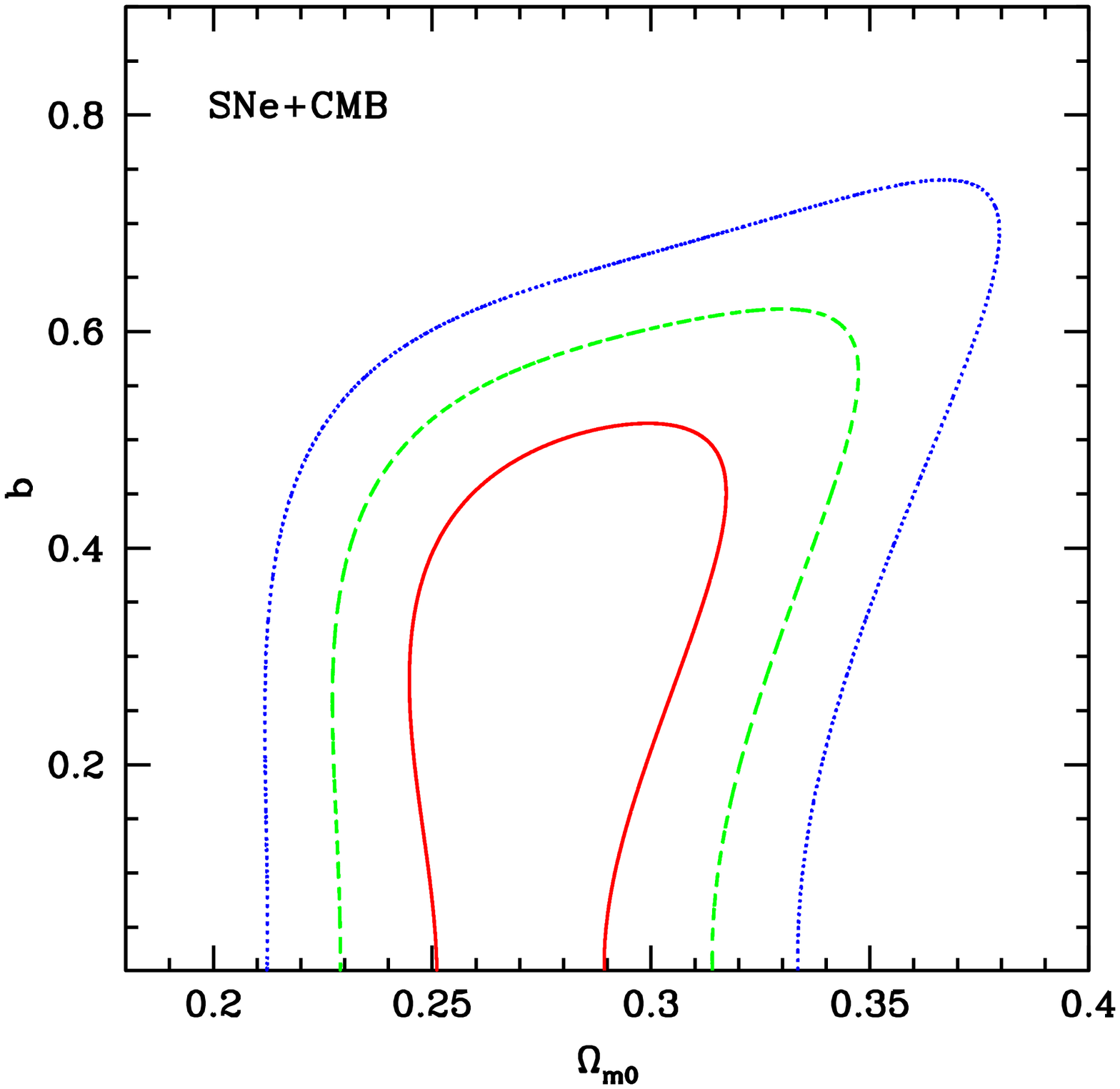,width=3.truein,height=3.truein,angle=0}}\\
\end{tabular}
\caption{The contour maps of $b-\Omega_{\rm m0}$ for different
sets. The left panel is $H(z)$+CMB, and the right panel is SNe
Ia+CMB. Confidence regions are at $68.3\%$, $95.4\%$ and $99.7\%$
levels from inner to outer respectively, for a flat Undulant
Universe without a prior of $h$.}\label{fig.3}
\end{figure*}

\begin{figure*}

\begin{tabular}{cc}

\parbox{58mm}{\psfig{figure=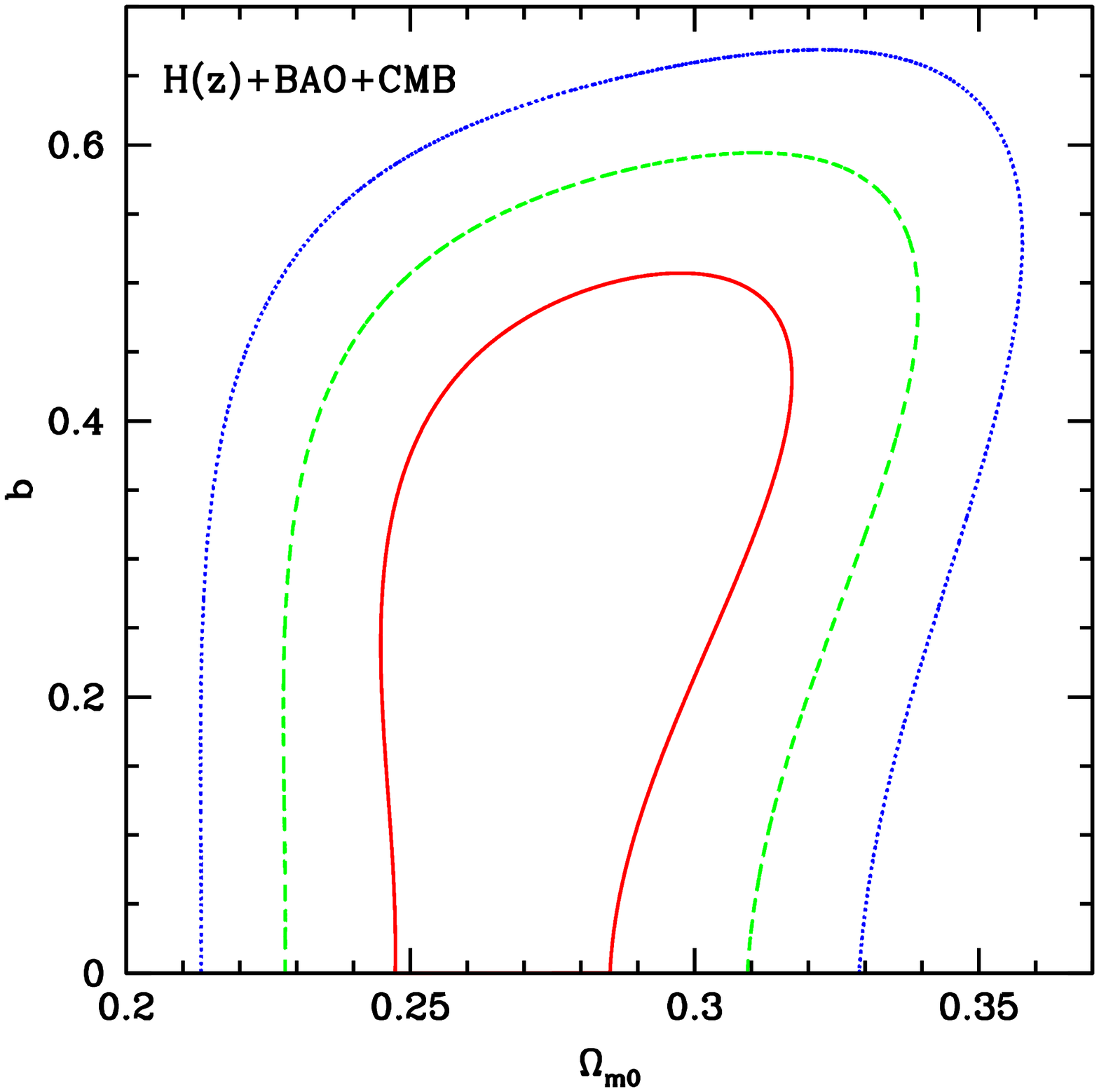,width=3.truein,height=3.truein,angle=0}}\
\parbox{58mm}{\psfig{figure=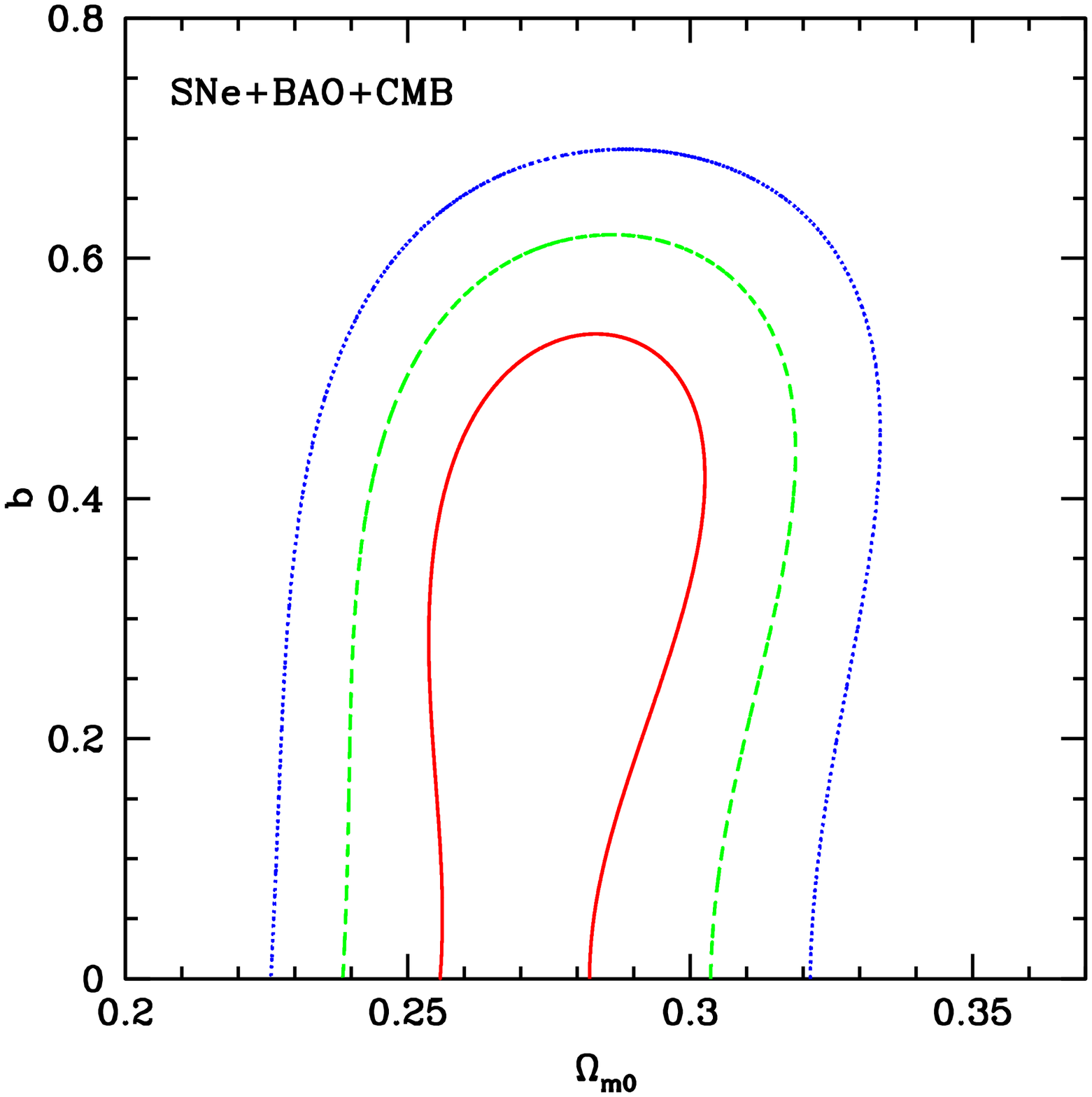,width=3.truein,height=3.truein,angle=0}}\\
\end{tabular}
\caption{The contour maps of $b-\Omega_{\rm m0}$ for different
sets. The left panel is $H(z)$+BAO+CMB, and the right panel is SNe
Ia+BAO+CMB. Confidence regions are at $68.3\%$, $95.4\%$ and
$99.7\%$ levels from inner to outer respectively, for a flat
Undulant Universe without a prior of $h$.}\label{fig.4}
\end{figure*}

\begin{figure*}

\begin{tabular}{cc}

\parbox{58mm}{\psfig{figure=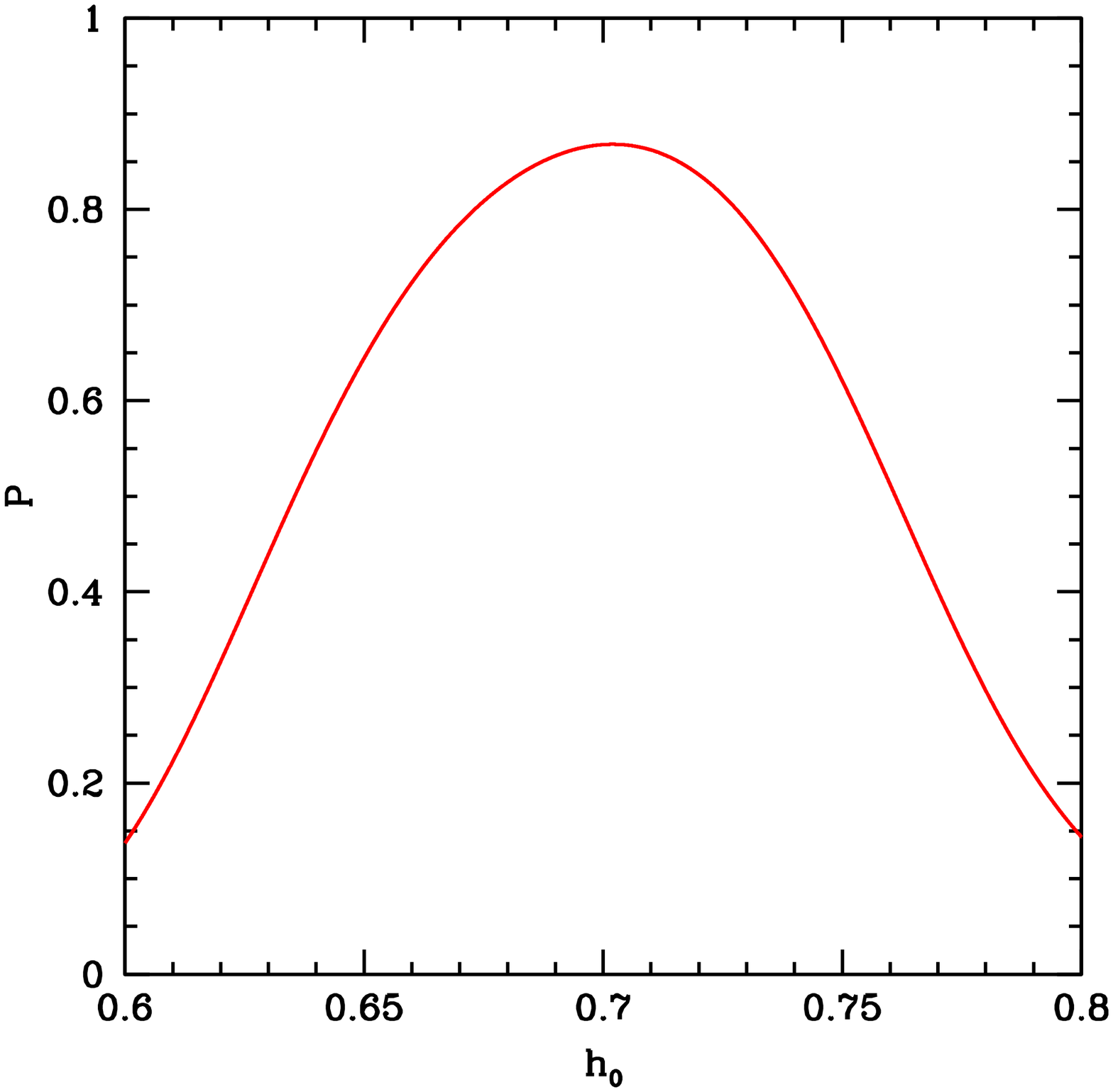,width=3.truein,height=3.truein,angle=0}}\
\parbox{58mm}{\psfig{figure=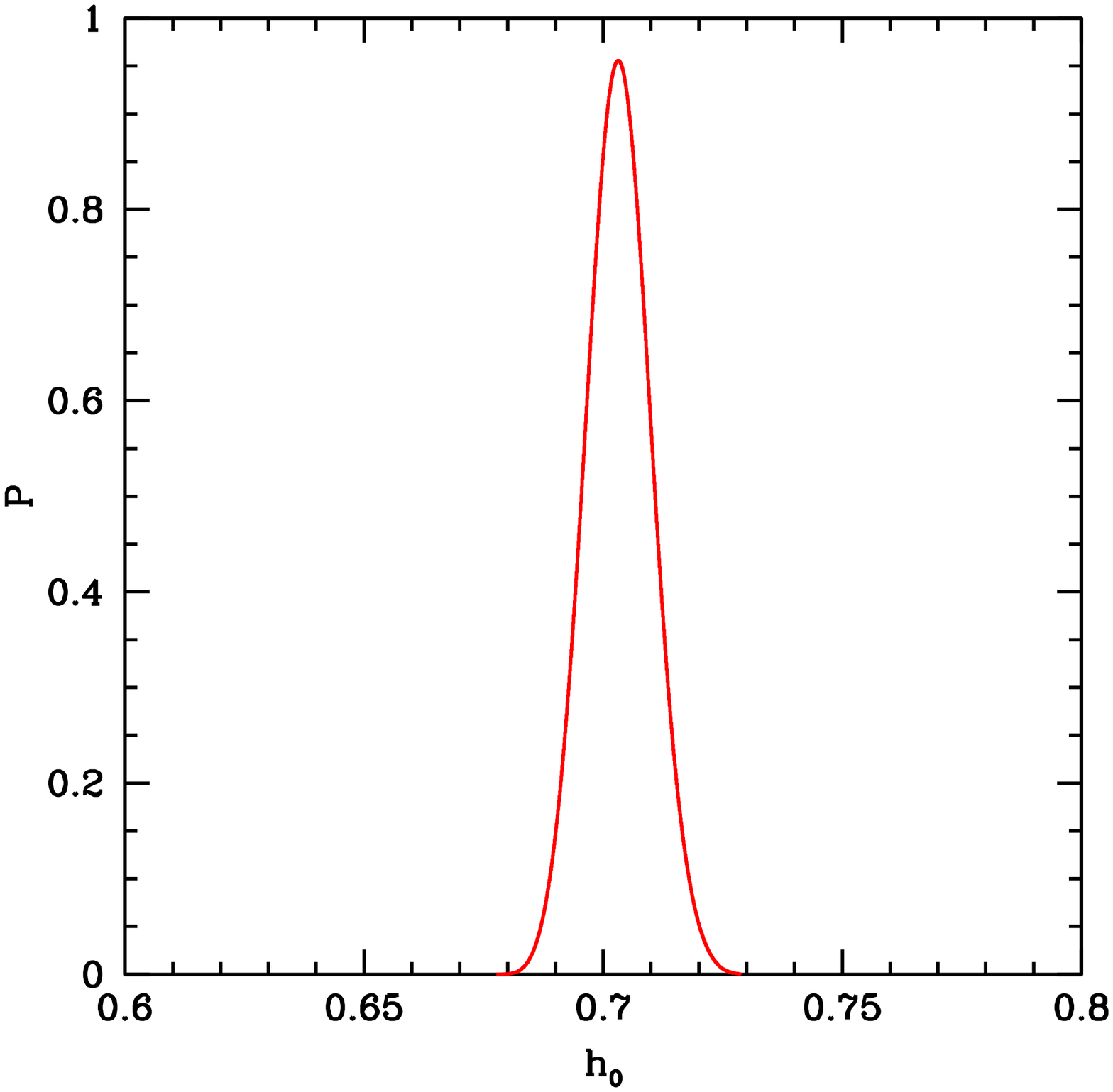,width=3.truein,height=3.truein,angle=0}}\\
\end{tabular}
\caption{The probability distribution function of $h$ for $H(z)$ and SNe Ia data, respectively, for a flat Undulant Universe.}\label{fig.5}
\end{figure*}

\begin{figure*}

\begin{tabular}{cc}

\parbox{58mm}{\psfig{figure=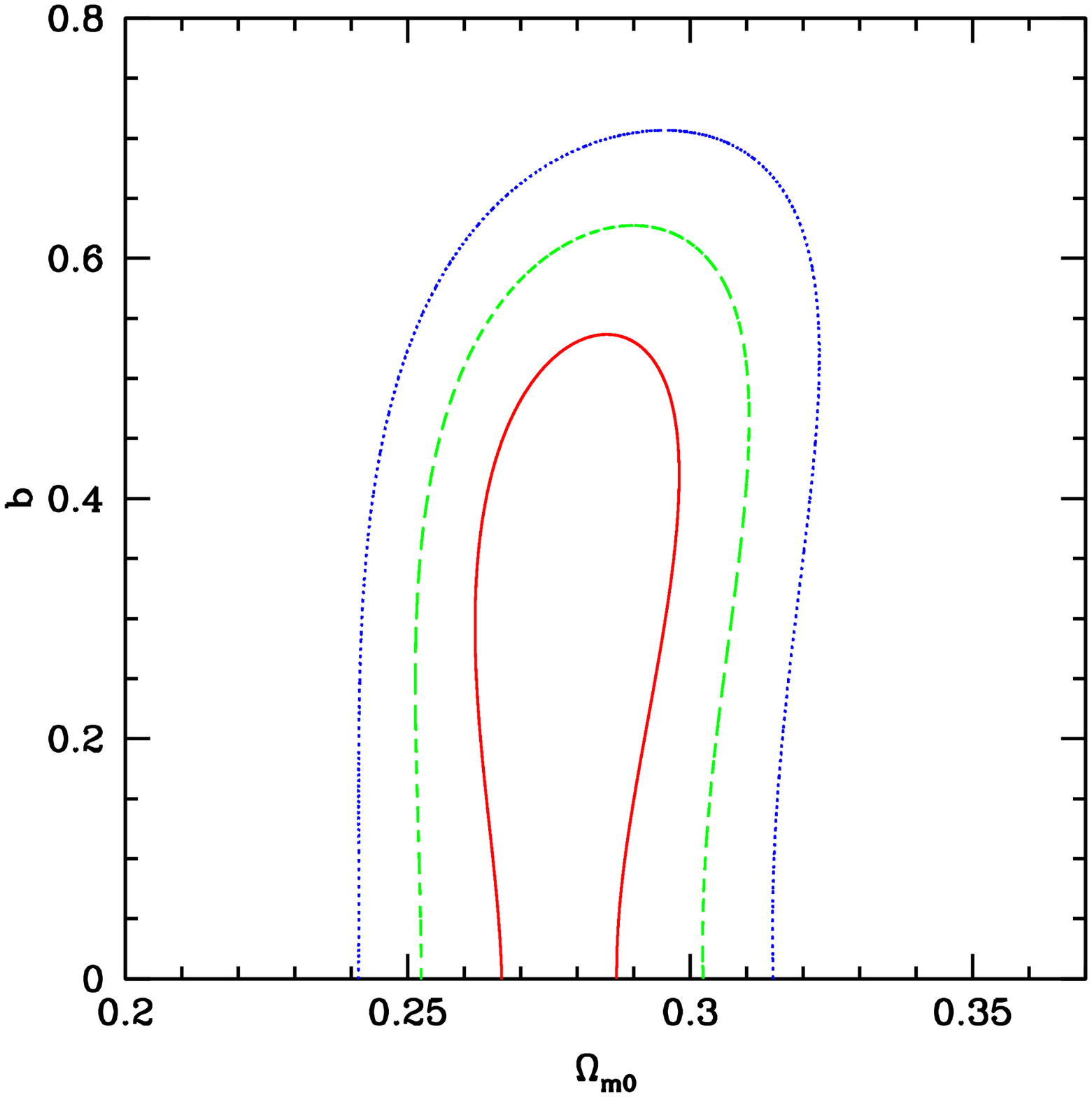,width=3.truein,height=3.truein,angle=0}}\
\parbox{58mm}{\psfig{figure=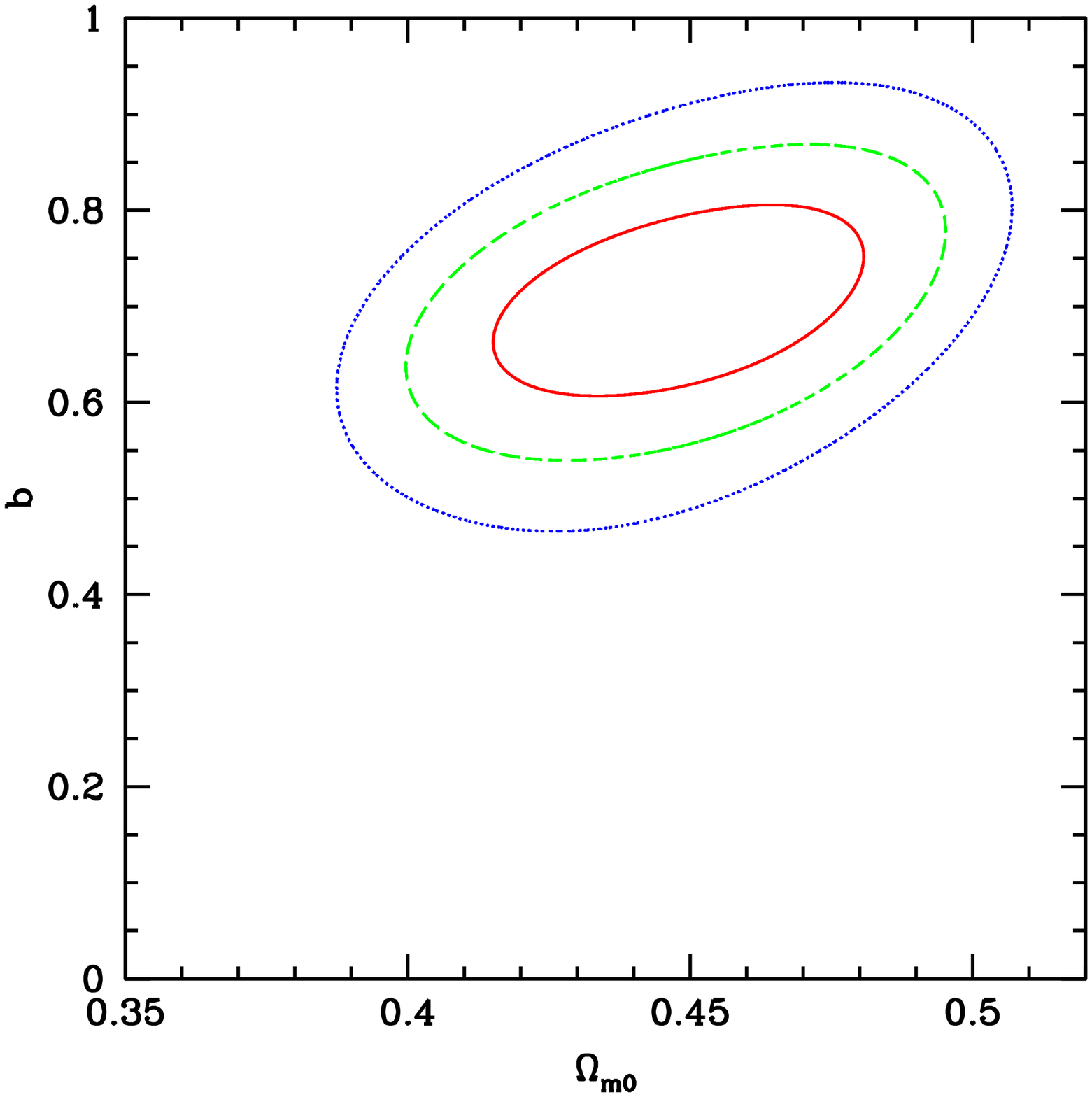,width=3.truein,height=3.truein,angle=0}}\\
\end{tabular}
\caption{The contour maps of $b-\Omega_{\rm m0}$ for SNe Ia+BAO+CMB with a "wrong" $h$ prior. $h$ is $0.7$ for the left panel, and $0.65$ for
the right panel. Confidence regions are at $68.3\%$, $95.4\%$ and $99.7\%$ levels from inner to outer respectively, for a flat Undulant Universe
without a prior of $h$.}\label{fig.6}
\end{figure*}

\section{Constraints on the Undulant Universe}
Considering that the contents of the universe are mainly dark
matter and dark energy, we just show $b-\Omega_{\rm m0}$ plane
contour maps, where $b$ traces dark energy, and $\Omega_{\rm m0}$
traces dark matter respectively. We estimate the best fit to the
set of parameters by using individual $\chi^2$ statistics, with
\begin{equation}
\chi^2_{\mid \rm data} = \sum^{\rm N}_{\rm i=1}\frac{[d^{\rm
i}_{\rm t}(z)-d^{\rm i}_{\rm o}(z)]^{2}}{\sigma^{2}_{\rm
i}},\label{eq3}
\end{equation}
where $\rm N$ is the number of data in each set, and $d^{\rm i}_{\rm t}(z)$ is the value at $z_{\rm i}$ given by above theoretical analysis.
$d^{\rm i}_{\rm o}(z)$ is individual observational value in each data set at $z_{\rm i}$, and $\sigma_{\rm i}$ is the error due to
uncertainties. In order to strengthen the constraints, we use MCMC technique to modulate the process (\cite{Gong2007}).

It is worth noting that the value of $h$ may strongly affect the constraint process, because the probability distribution function (PDF) of $h$
for each data is different. We can see the PDF for $H(z)$ and SNe Ia data in Fig.5. For $H(z)$ data, we find the PDF of $h$ covers a large
range, while it is small for SNe Ia date. Because the distribution of $h$ constrained by SNe Ia is very narrow, if we fix $h$ far from the
best-fit, then the constraints will be totally wrong.For example, we test the effect on fixing a "wrong" $h$ by using SNe Ia+BAO+CMB. As can be
seen in Fig.6, with a "wrong" $h$ prior, the constraints of the left panel are very different from that of the right panel. This is because if
we assume a "wrong" prior for $h$, the $\chi^2$ would become very large ($\chi^2\sim426$ for $h=0.65$ while $\chi^2\sim312$ for $h=0.7$), and
then we will get a false minimum $\chi^2$ to lead to a wrong result.However, if we marginalize the likelihood functions over $h$ by integrating
the probability density $P\propto e^{-\chi^2/2}$, we can remove the effect of the distribution of $h$, and illustrate the reliable distribution
of the other parameters. Therefore, we marginalize the likelihood functions over $h$ to obtain the best fitting results. Then we get the
confidence regions and the $b-\Omega_{\rm m0}$ plane.

As it is seen from Eq.(1), $\omega(a)$ is an even function. Correspondingly, the values of $b$ are along the axis of $\Omega_{\rm m0}$
symmetrical distribution. Thus, the constraints on $b$ just in the range from $0$ to $+\infty$ are enough to demonstrate the comparison. Fig.1
shows the confidence regions determined by the $H(z)$ and SNe Ia data alone respectively. We find that the constraints on $\Omega_{\rm m0}$ from
the $H(z)$ data are weaker than that from the SNe Ia data. Although we just have 9 data in $H(z)$ and 307 data in SNe Ia, there is still some
consistency between them in the constraints. The joint constraints are shown in Fig.2, Fig.3, and Fig.4. We find that the constraints are weak
if we use the $H(z)$ or SNe Ia data alone, however, after inclusion of the other observational data (the BAO or CMB data), the constraints are
improved remarkably. Especially, the CMB data improve the constraints more remarkably. For breaking the degeneracy, the BAO and CMB data play
very important roles in the constraint compared with the $H(z)$ and SNe Ia data alone. In the left contour map of Fig.1, and the two contour
maps of Fig.2, Fig.3, and Fig.4, all contours at the $1 \sigma$ confidence for $b$ are open-ended. It shows that $b$ takes possible value around
$0$. We have not demonstrated the details of the constraints, because MCMC technique cannot determine exact values reliably. These contours
correspond to $1 \sigma, 2 \sigma$, and $3 \sigma$ confidence from inside to outside. Table.\ref{tb2} summarizes $68.3\%$ confidence intervals.
From the table, we find that the constraints on $\Omega_{\rm m0}$ from $H(z)$+BAO are looser than that from SNe Ia+BAO, and there is little
difference between them. However, the constraints on $b$ from SNe Ia+BAO are looser than that from $H(z)$+BAO. The similar conclusions can be
derived from comparing $H(z)$+CMB with SNe Ia+CMB, and $H(z)$+BAO+CMB with SNe Ia+BAO+CMB. The results of $\Omega_{\rm m0}$ on the Undulant
Universe are consistent with the recent reported results from WMAP 5-year data (\cite{Kowalski2008}). The distribution of $b$ is close to $0$,
i.e. the Undulant Universe approaches the $\Lambda \rm$CDM model. Comparing Fig.4 with Fig.6, we find that the result of the right panel in
Fig.6 with $h=0.7$ is similar with that of the right panel in Fig.4 with marginalizing $h$, but has a smaller range for $\Omega_{\rm m0}$. On
the other hand, from the result of the left panel in Fig.6 with $h=0.65$, we find that the range of $b$ shrinks to $0.6\sim0.8$ at $68.3\%$ C.L.
and the best-fit value of $\Omega_{\rm m0}$ moves to $\sim0.45$ around, which are wrong results.

\begin{table}
\caption{$68.3\%$ confidence intervals for each observational
data}
\begin{tabular}{ccccccccc}
\noalign{\smallskip} \hline \noalign{\smallskip}
Data & $H(z)$ & SNe & $H(z)$+BAO & SNe+BAO \\
\hline
$\Omega_{m0}$&$[0.158,0.43]$&$[0.150,0.299]$&$[0.239,0.315]$&$[0.239,0.305]$\\
$b$          &$[0,2.0]$&$[1.13,3.21]$&$[0,1.95]$&$[0,2.47]$\\
\noalign{\smallskip} \hline \noalign{\smallskip}
\noalign{\smallskip} \hline \noalign{\smallskip}
Data & $H(z)$+CMB & SNe+CMB & $H(z)$+BAO+CMB & SNe+BAO+CMB \\
\hline
$\Omega_{m0}$&$[0.21,0.30]$&$[0.244,0.318]$&$[0.244,0.318]$&$[0.253,0.304]$\\
$b$          &$[0,0.40]$&$[0,0.53]$&$[0,0.51]$&$[0,0.58]$\\
\noalign{\smallskip} \hline \noalign{\smallskip}
\end{tabular}
\label{tb2}
\end{table}

\section{Conclusions and Discussions}
So far, we've presented some kinds of constraints on cosmological parameters by using the $H(z)$, SNe Ia, BAO and CMB data on the flat Undulant
Universe, and we have marginalized the parameter $h$ to get the best fitting results. Because if we fix a "wrong" prior for $h$, any model we
take cannot fit the observational data, and then leads to a false $\chi^2$. We show confidence regions on the $b-\Omega_{\rm m0}$ plane. The
Undulant model can trace the distribution of dark energy in undulant case. Therefore, our conclusions are more general. As discussed above, we
have showed the comparison of different data sets carefully. Comparing with the $H(z)$ and SNe Ia data alone, the BAO and CMB data play very
important roles in breaking the degeneracy. The constraints on $\Omega_{\rm m0}$ from the $H(z)$ data are looser than that from the SNe Ia data,
but there is little difference between them. For the combined data sets, the constraints are improved remarkably. The CMB data improve the
constraints more remarkably than the BAO data. In the joint constraints, the results from the $H(z)$ data sets ($H(z)$+BAO, $H(z)$+CMB and
$H(z)$+BAO+CMB) are almost consistent with that from the SNe Ia data sets (SNe Ia+BAO, SNe Ia+CMB and SNe Ia+BAO+CMB). We find all the
constraints on $b$ are around $0$, but the $H(z)$ data sets constrain $b$ more tightly. Namely, the Undulant Universe approaches the $\Lambda
\rm$CDM model. However, on $\Omega_{\rm m0}$, the constraints of the $H(z)$ data sets are weaker than that of the SNe Ia data sets. It is
probably because the data amount is not sufficient and the corresponding errors are very large (\cite{Samushia2006}). Only are there 9 $H(z)$
data in our discussion, but we still get such fine tight constraints. We speculate that the constraints of the $H(z)$ data sets will be tighter
if we get more $H(z)$ data. Therefore, much more $H(z)$ data have an advantage over the SNe Ia data in the constraint. Fortunately, a large
amount of $H(z)$ including data from the AGN and Galaxy Evolution Survey (AGES) and the Atacama Cosmology Telescope (ACT) are expected to be
available in the next few years (\cite{Samushia2006}). Provided a statistical sample of many hundreds of galaxies, we could determine the value
of $H(z)$ data to a percent accuracy. Paying attention to the assumption that SNe Ia is ``standard candle'', it might make the luminosity
distance $d_L$ of SNe Ia sample biased. In addition, there exists an integration of the inverse of $H(z)$ in $d_L$, and some uncertainties in
the minimizing $\chi^2$ statistics might also arise from this integration. On the contrary, $H(z)$ do not suffer from this uncertainty resulting
from integration. Therefore, the $H(z)$ data should constrain parameters more strongly than the SNe Ia data. We also notice that, the redshift
range of $H(z)$ is from 0.09 to 1.75 and the redshift range of SNe Ia is from about 0.01 to 1.75. Thus, $H(z)$ and SNe Ia are distributed in the
same range of redshift roughly. Therefore, the $H(z)$ data seem to provide us with an independent, very simple, and very powerful probe of
fundamental cosmological parameters. With a large amount of the $H(z)$ data in the future, we probably can constrain cosmological parameters by
using the $H(z)$ data instead of the SNe Ia data. \normalem

\begin{acknowledgements}
We are very grateful to the anonymous referee for many valuable comments that greatly improved the paper. Tian Lan would like to thank Hui Lin
for her valuable discussions, and L. Shao for his help in English. Our MCMC chain computation was performed on the Supercomputing Center of the
Chinese Academy of Sciences and the Shanghai Supercomputing Center. This work was supported by the National Science Foundation of China (Grants
No.10473002), the Ministry of Science and Technology National Basic Science program (project 973) under grant No.2009CB24901, Scientific
Research Foundation of Beijing Normal University and the Scientific Research Foundation for the Returned Overseas Chinese Scholars, State
Education Ministry.
\end{acknowledgements}

\end{document}